\documentclass[a4paper,11pt]{article}
\usepackage{pos} 

\usepackage{tikz}
\usetikzlibrary{shapes.geometric, arrows.meta, positioning}

\tikzstyle{box} = [rectangle, rounded corners, minimum width=30mm, minimum height=10mm,text centered, draw=black, fill=white]
\tikzstyle{obs} = [rectangle, rounded corners, minimum width=3mm, minimum height=3mm,text centered, draw=black, fill=red!20]
\tikzstyle{arrow} = [red,line width=1.5pt,-{Stealth[length=3mm]},shorten >=3pt]

\usepackage{ifthen}
\newboolean{articletitles}
\setboolean{articletitles}{true}
\newboolean{inbibliography}
\setboolean{inbibliography}{false}
\usepackage{cite}
\usepackage{mciteplus}

\title{Taming Penguins: Towards High Precision Measurements in $\phi_d$ and $\phi_s$}
\ShortTitle{Taming Penguins}

\author*[a,b]{Kristof De Bruyn}
\author[a,c]{Robert Fleischer}
\author[d]{Eleftheria Malami}

\affiliation[a]{Nikhef,\\
  Science Park 105, 1098 XG Amsterdam, Netherlands}

\affiliation[b]{Van Swinderen Institute for Particle Physics and Gravity, University of Groningen,\\
9747 Groningen, Netherlands}

\affiliation[c]{Faculty of Science, Vrije Universiteit Amsterdam,\\
1081 HV Amsterdam, Netherlands}

\affiliation[d]{Center for Particle Physics Siegen, University of Siegen,\\
D-57068 Siegen, Germany}

\emailAdd{k.a.m.de.bruyn@rug.nl}

\abstract{
Experimentally, the phases $\phi_d$ and $\phi_s$ are determined from CP asymmetry measurements in the ``golden modes'' $B_d^0\to J/\psi K_{\mathrm{S}}^0$ and $B_s^0\to J/\psi\phi$. 
At leading order, the theoretical interpretation of these measurements is straightforward.
However, to reach high precision determinations of $\phi_d$ and $\phi_s$, which is essential in view of the searches for signs of beyond the SM physics, corrections from next-to-leading order effects need to be accounted for.
These corrections primarily originate from so-called penguin topologies.
Using the $SU(3)$ flavour symmetry, these corrections can be determined using suitably chosen control modes.
Recent new CP asymmetry measurements from LHCb in $B\to DD$ and Belle-II in $B_d^0\to J/\psi\pi^0$ decays greatly improve our knowledge on the parameters describing the contribution from penguin topologies.
These proceedings will discuss the current constraints on the penguin parameters in $B\to J/\psi X$ and $B\to DD$ decays, provide corrected determinations for $\phi_d$ and $\phi_s$, and highlight what can be expected at the end of the HL-LHC and Belle-II programmes.
}

\FullConference{
9th Symposium on Prospects in the Physics of Discrete Symmetries (DISCRETE 2024)\\
2-6 December, 2024\\
Ljubljana
}

\begin{document}
\maketitle

\section{Introduction}
In the Standard Model (SM), the complex phases $\phi_d$ and $\phi_s$ associated with the mixing between neutral $B_q^0$ and $\bar B_q^0$ mesons, where $q\in\{d,s\}$, are parametrised as
\begin{equation}\label{eq:phi_SM}
    \phi_d^{\text{SM}} = 2\beta\:, \qquad
    \phi_s^{\text{SM}} = 2\lambda^2\eta\:.
\end{equation}
Their experimental determinations are powerful probes to search for evidence of physics beyond the Standard Model (BSM):
The measurement of $\phi_d$ constrains the angle $\beta$ of the Unitarity Triangle (UT), and the value of $\phi_s$ can directly be compared with the SM prediction based on the Wolfenstein parametrisation \cite{Wolfenstein:1983yz,Buras:1994ec} of the Cabibbo--Kobayashi--Maskawa (CKM) quark-mixing matrix \cite{Cabibbo:1963yz,Kobayashi:1973fv}.

The phases $\phi_q$ are experimentally determined from Charge-Parity (CP) asymmetries created by the interference between the $B_q^0$--$\bar B_q^0$ mixing process and the decay of the $B_q^0$ or $\bar B_q^0$ meson.
The most precise measurements of $\phi_d$ and $\phi_s$ use respectively the ``golden'' decays $B_d^0\to J/\psi K_{\mathrm{S}}^0$ and $B_s^0\to J/\psi\phi$.
The decay processes of these two channels are dominated by a colour-suppressed tree topology.
When only considering this decay topology and ignoring other, sub-leading contributions, the mixing-induced CP asymmetry becomes proportional to $\sin(\phi_q)$.
The uncertainty associated with ignoring the sub-leading contributions, which are primarily due to penguin topologies, is as large as $0.5^{\circ}$ \cite{Barel:2020jvf}.
Given the experimental precision that already has been achieved today, and looking at the prospects for the end of the Belle-II \cite{Belle-II:2018jsg} and High Luminosity LHC (HL-LHC) \cite{LHCb:2018roe} programmes, this will become the leading systematic uncertainty and limit our potential to search for BSM physics using $\phi_d$ and $\phi_s$.
When including sub-leading effects due to penguin topologies, the CP asymmetries allow us to determine an effective mixing phase
\begin{equation}\label{eq:eff_mix_phase}
    \phi_{q}^{\text{eff}} \equiv \phi_q + \Delta\phi_q = \phi_q^{\text{SM}} + \phi_q^{\text{NP}} + \Delta\phi_q\:,
\end{equation}
where $\phi_q^{\text{NP}}$ is the potential new physics (NP) contribution from BSM physics that we wish to constrain, and $\Delta\phi_q$ is a decay-channel-specific hadronic phase shift due to the sub-leading SM processes.
The current experimental picture suggests that $\phi_q^{\text{NP}}$ and $\Delta\phi_q$ can be similar in size, i.e.\ zero or small, thereby making it challenging to disentangle them from each other.
High precision measurements of $\phi_{q}^{\text{eff}}$ thus need to be complemented with accurate knowledge on the phase shift $\Delta\phi_q$.

These proceedings present results on the phase shift $\Delta\phi_q$ affecting the measurements of $\phi_d$ and $\phi_s$ in the decay channels $B_d^0\to J/\psi K_{\mathrm{S}}^0$, $B_s^0\to J/\psi\phi$ and $B_s^0\to D_s^+D_s^-$.
The impact of the penguin topologies is determined using the strategy originally proposed in Refs.\ \cite{Fleischer:1999nz,Fleischer:1999zi}, which relies on the $SU(3)$ flavour symmetry of QCD.
For each of the three above decays a control mode is identified in which the penguin topologies are enhanced compared to the leading tree topology.
The measured CP asymmetries of the control modes allow us to quantify the contributions from the penguin topologies, which, assuming the $SU(3)$ flavour symmetry, can be related to the penguin contributions in $B_d^0\to J/\psi K_{\mathrm{S}}^0$, $B_s^0\to J/\psi\phi$ and $B_s^0\to D_s^+D_s^-$.
For $B_d^0\to J/\psi K_{\mathrm{S}}^0$, our chosen control modes are $B_s^0\to J/\psi K_{\mathrm{S}}^0$ and $B_d^0\to J/\psi\pi^0$, while for $B_s^0\to J/\psi\phi$ the primary control mode is $B_d^0\to J/\psi\rho^0$, and for $B_s^0\to D_s^+D_s^-$ it is $B_d^0\to D^+D^-$.
The Belle-II collaboration has recently released new measurements of the CP asymmetries in $B_d^0\to J/\psi\pi^0$ \cite{Belle-II:2024hqw}, and the LHCb collaboration provided new results on the CP asymmetries in $B_d^0\to D^+D^-$ \cite{LHCb:2024gkk}.
In both cases, the new measurements have a precision which is similar to the old world average compiled by the HFLAV collaboration \cite{HFLAV:2022pwe} and will thus significantly improve the constraints on the penguin effects.

\section{Fit for the penguin parameters}
The impact on the determination of $\phi_d$ and $\phi_s$ from the penguin topologies in the $B_d^0\to J/\psi K_{\mathrm{S}}^0$, $B_s^0\to J/\psi\phi$ and $B_s^0\to D_s^+D_s^-$ decays is determined following the framework described in Refs. \cite{Bel:2015wha,Barel:2020jvf}.
For each of the three decays, the decay amplitude is parametrised as
\begin{equation}\label{eq:amplitude}
    A(B_q^0\to f) = \mathcal{N}\left(1 + \epsilon a e^{i\theta}e^{i\gamma}\right)\:,
\end{equation}
where $\epsilon \approx 0.052$, $a$ parametrises the relative size of the penguin topologies compared to the leading tree topology, $\theta$ is the associated strong phase difference, and the UT angle $\gamma$ gives the weak phase difference between the topologies.
$\mathcal{N}$ is a normalisation factor which drops out in the CP asymmetries.
The penguin parameters $a$ and $\theta$ depend on the decay dynamics and can thus be different for the three decays.
We therefore use $a_{J/\psi P}$ to parametrise $B_d^0\to J/\psi K_{\mathrm{S}}^0$ and its control modes, $a_{J/\psi V}$ for $B_s^0\to J/\psi\phi$ and $a_{DD}$ for $B_s^0\to D_s^+D_s^-$.

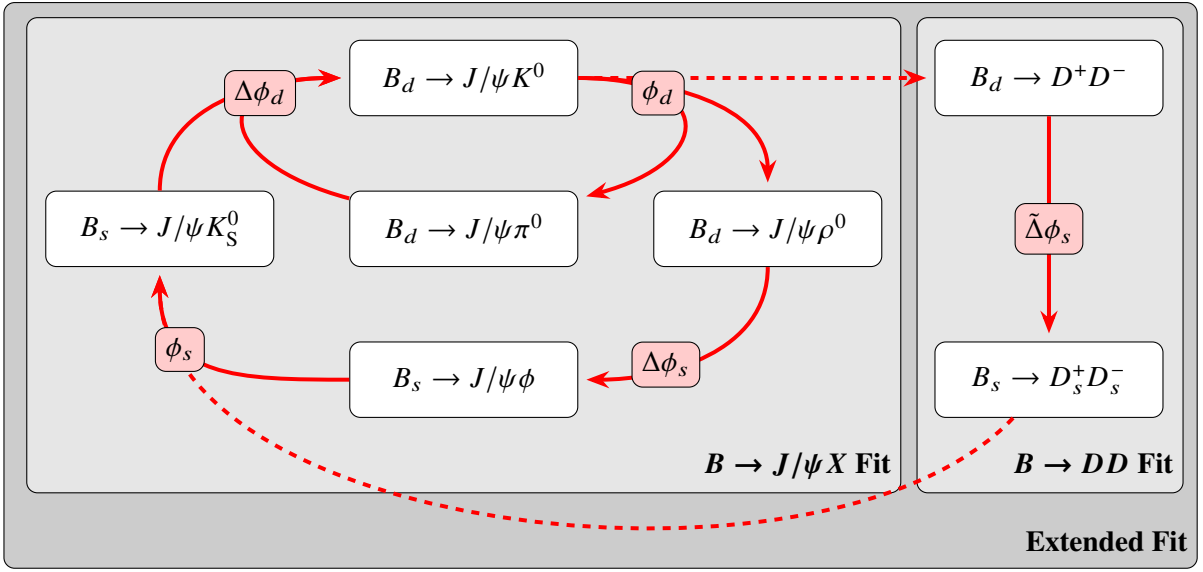
\begin{figure}
    \center
\begin{tikzpicture}[node distance=20mm]
    \node (fitAll) [draw=black, fill=black!20, minimum width=157mm,minimum height=75mm, rounded corners]{};
    \node (fitJX) [left=18mm of fitAll.center, anchor=center, yshift=4mm, draw=black, fill=black!10, minimum width=115mm,minimum height=63mm, rounded corners]{};
    \node (fitDD) [right=2mm of fitJX, draw=black, fill=black!10, minimum width=35mm,minimum height=63mm, rounded corners]{};
    
    \node (textJX) [above=0.5mm of fitJX.south east, anchor=south east, font=\boldmath\bf] {$B\to J/\psi X$ Fit};
    \node (textDD) [above=0.5mm of fitDD.south east, anchor=south east, font=\boldmath\bf] {$B\to DD\vphantom{\psi}$ Fit};
    \node (textAll) [above=1mm of fitAll.south east, anchor=south east, font=\boldmath\bf] {Extended Fit};

    \node (BdJpsiK) [box, below=3mm of fitJX.north, anchor=north] {$B_d\to J/\psi K^0$};
    \node (BdJpsiPi) [box, below of=BdJpsiK] {$B_d\to J/\psi \pi^0$};
    \node (BdJpsiRho) [box, right of=BdJpsiPi, xshift=20mm] {$B_d\to J/\psi \rho^0$};
    \node (BsJpsiK) [box, left of=BdJpsiPi, xshift=-20mm] {$B_s\to J/\psi K_{\text{S}}^0$};
    \node (BsJpsiPhi) [box, below of=BdJpsiPi] {$B_s\to J/\psi \phi$};
    \node (BdDD) [box, right of=BdJpsiK, xshift=57mm] {$B_d\to D^+D^-$};
    \node (BsDsDs) [box, right of=BsJpsiPhi, xshift=57mm] {$B_s\to D_s^+D_s^-$};

    \draw[arrow] (BdJpsiK) to [out=0,in=90] (BdJpsiRho);
    \draw[arrow] (BdJpsiK) to [out=0,in=15,looseness=3] (BdJpsiPi);
    \draw[arrow, dashed] (BdJpsiK) to [out=0,in=180] (BdDD);

    \draw[arrow] (BdJpsiRho) to [out=270,in=0] (BsJpsiPhi);
    \draw[arrow] (BdDD) to [out=270,in=90] (BsDsDs);

    \draw[arrow] (BsJpsiPhi) to [out=180,in=270, looseness=1.3] (BsJpsiK);
    \draw[arrow, dashed] (BsDsDs) to [out=227,in=270, looseness=0.8] (BsJpsiK);

    \draw[arrow] (BsJpsiK) to [out=90,in=180] (BdJpsiK);
    \draw[arrow] (BdJpsiPi) to [out=165,in=180,looseness=3] (BdJpsiK);

    \node (phid) [obs, right=7mm of BdJpsiK, yshift=-2mm] {$\phi_d$};
    \node (phis) [obs, left=19mm of BsJpsiPhi, yshift=4mm] {$\phi_s$};
    \node (Dphid) [obs, left=7mm of BdJpsiK, yshift=-2mm] {$\Delta\phi_d$};
    \node (Dphis) [obs, right=7mm of BsJpsiPhi, yshift=2mm] {$\Delta\phi_s$};
    \node (Dphis2) [obs, below of=BdDD] {$\tilde\Delta\phi_s$};
\end{tikzpicture}
    \caption{Schematic overview of the seven considered decay channels and their interdependence.}
    \label{fig:models}
\end{figure}

The mixing-induced CP asymmetries in the control modes also depend on the mixing phases $\phi_d$ and $\phi_s$.
To take this interdependence, illustrated in Fig.\ \ref{fig:models}, into account, we propose a simultaneous fit to the direct and mixing-induced CP asymmetries in all seven decay channels.
The UT angle $\gamma = (65.6_{-3.0}^{+2.9})^{\circ}$ \cite{HFLAV:2022pwe} is taken as external input.
The fit is performed using the GammaCombo framework \cite{LHCb:2016mag}.

\begin{figure}
    \centering
    \includegraphics[width=0.47\textwidth]{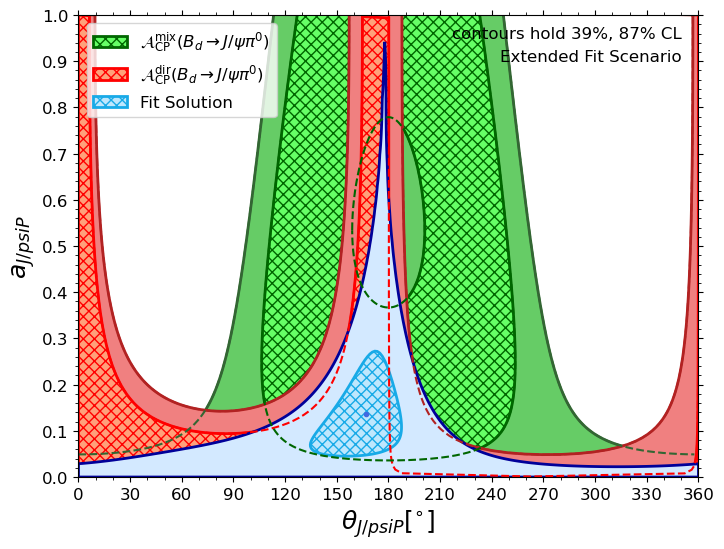}
    \includegraphics[width=0.47\textwidth]{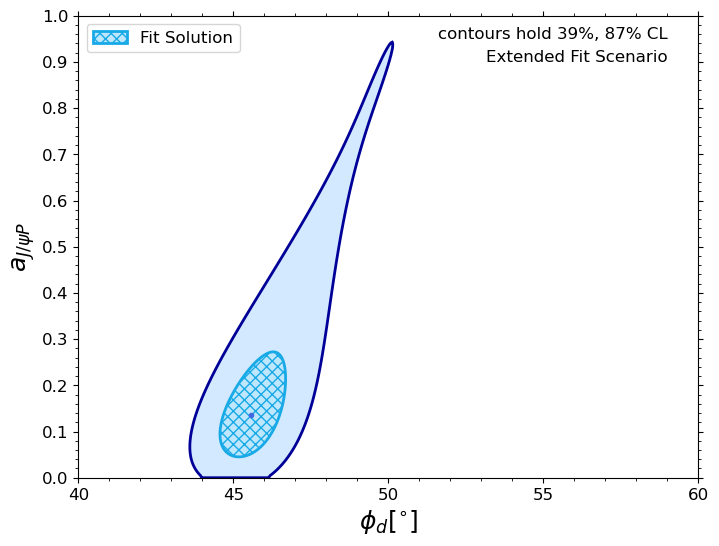}

    \includegraphics[width=0.47\textwidth]{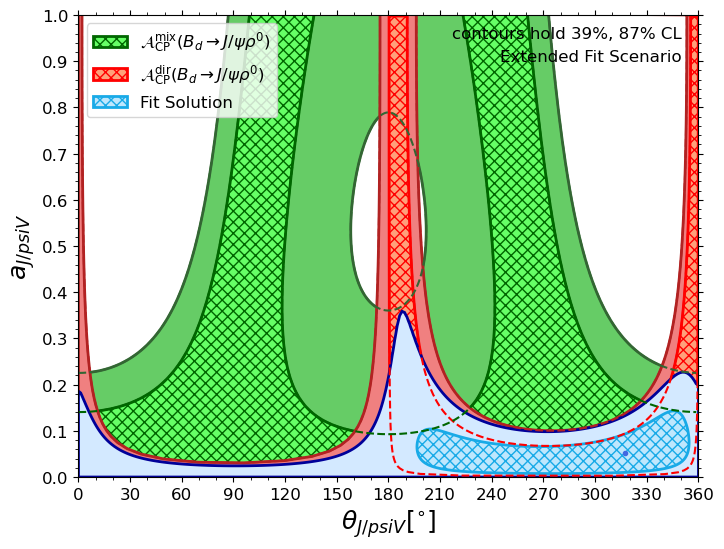}
    \includegraphics[width=0.47\textwidth]{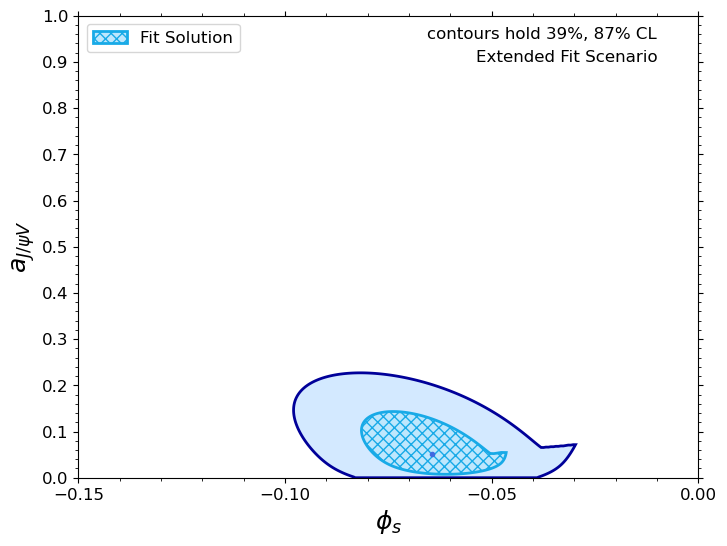}

    \includegraphics[width=0.47\textwidth]{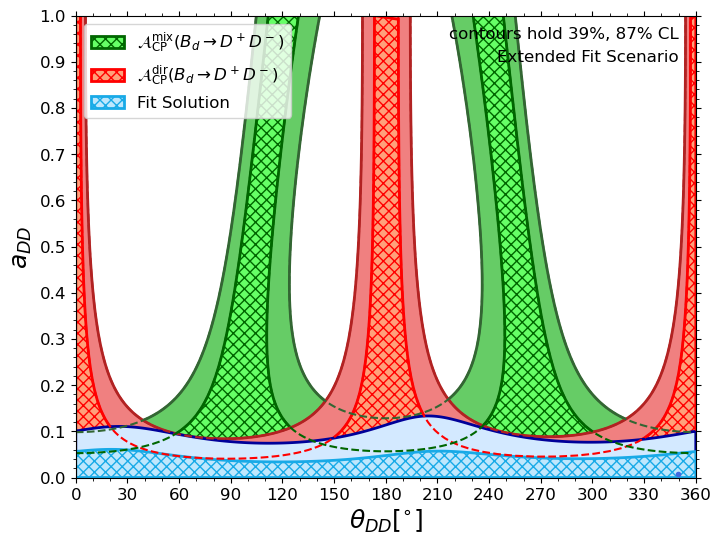}
    \includegraphics[width=0.47\textwidth]{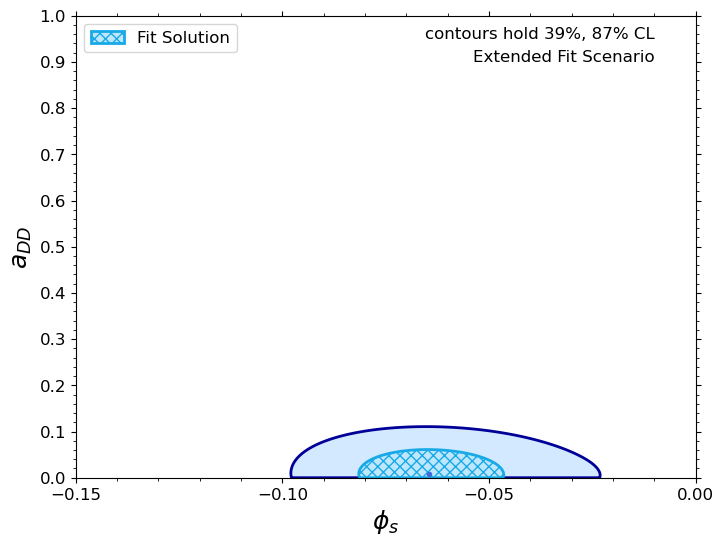}
    \caption{Two-dimensional confidence regions of the fit for the penguin parameters, $\phi_d$ and $\phi_s$ from the measured CP asymmetries in the $B\to J/\psi X$ and $B\to DD$ decays.
    Note that the contours for $\mathcal{A}_{\text{CP}}^{\text{dir}}$ and $\mathcal{A}_{\text{CP}}^{\text{mix}}$ are added for illustration only.
    They include the best fit solutions for $\phi_d$, $\phi_s$ and $\gamma$ as Gaussian constraints.
    }
    \label{fig:Autumn24_ext}
\end{figure}

The extended fit results in the following values for the hadronic parameters:
\begin{align}
    a_{J/\psi P} & = 0.14_{-0.09}^{+0.14}\:, &
    \theta_{J/\psi P} & = \left(167_{-32}^{+21}\right)^{\circ}\:, \label{eq:extended_fit_I}\\
    a_{J/\psi V} & = 0.052_{-0.045}^{+0.092}\:, &
    \theta_{J/\psi V} & = \left(317_{-120}^{+38}\right)^{\circ}\:, \\
    a_{DD} & = 0.007_{-0.007}^{+0.054}\:, &
    \theta_{DD} & = \left(350_{-350}^{+10}\right)^{\circ}\:.
\end{align}
Taking these penguin effects into account, the state-of-the-art values for the mixing phases are
\begin{equation}\label{eq:extended_fit_IV}
    \phi_d = \left(45.6_{-1.0}^{+1.1}\right)^{\circ}\:,
    \qquad
    \phi_s = -0.065_{-0.017}^{+0.019} = \left(-3.72_{-0.97}^{+1.09}\right)^{\circ}\:,
\end{equation}
which can be compared to the effective mixing phases measured in $B_d\to J/\psi K^0$ and $B_s\to J/\psi \phi$:
\begin{equation}
    \phi_d^{\text{eff}} = \left(45.12 \pm 0.94\right)^{\circ}\:,
    \qquad
    \phi_s^{\text{eff}} = -0.061 \pm 0.014 = \left(3.50\pm 0.80\right)^{\circ}\:.
\end{equation}
The two-dimensional confidence level contours for the penguin parameters and mixing phases are shown in Fig. \ref{fig:Autumn24_ext}.

\section{\boldmath $SU(3)$-symmetry breaking}
\begin{figure}
    \centering
    \includegraphics[width=0.47\textwidth]{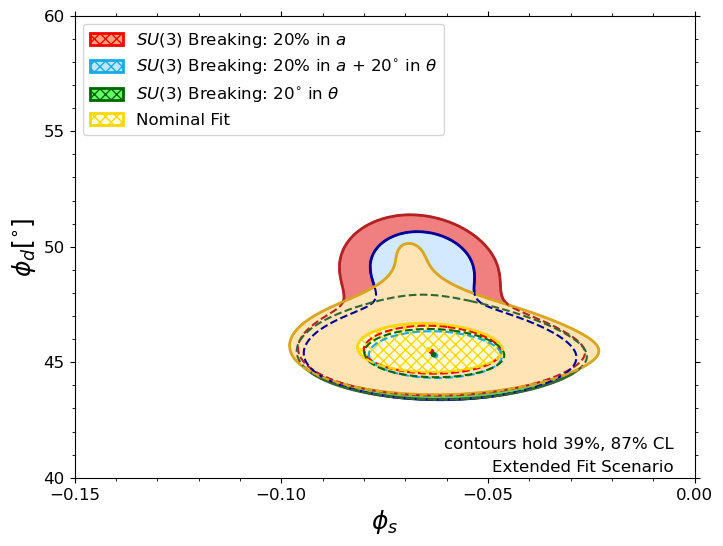}
    \caption{Two-dimensional confidence regions of the fit for $\phi_d$ and $\phi_s$ from the measured CP asymmetries in the $B\to J/\psi X$ and $B\to DD$ decays, assuming different $SU(3)$-breaking scenarios.
    The $SU(3)$-breaking effects are included as Gaussian constraints in the fit, and not free parameters in the model.
    }
    \label{fig:phi_SU3}
\end{figure}

The main systematic uncertainty associated with our method to control penguin contributions in the determination of $\phi_d$ and $\phi_s$ comes from potential $SU(3)$-breaking effects.
Because the CP asymmetries are ratios of decay amplitudes, any factorisable $SU(3)$-breaking effects, which impact the normalisation $\mathcal{N}$ in Eq.\ \eqref{eq:amplitude}, necessarily drop out.
This only leaves non-factorisable $SU(3)$-breaking effects, which are, in general, assumed to be subdominant compared to their factorisable counterparts.
The $SU(3)$-breaking effects can be introduced in the fit framework as external correction factors.
In Fig.\ \ref{fig:phi_SU3}, we compare the nominal fit to three hypothetical scenarios in which we consider correction factors $x_{SU(3)} = 1.2 \pm 0.2$ to the three penguin parameters $a$, and/or correction factors $y_{SU(3)} = (20 \pm 20)^{\circ}$ to the three penguin parameters $\theta$.
There is almost no impact on the inner contour, corresponding to the $1\sigma$ uncertainties on $\phi_d$ and $\phi_s$.
These results are used to assign a systematic uncertainty of $0.3^{\circ}$ to $\phi_d$ and 2 mrad $\left(0.11^{\circ}\right)$ to $\phi_s$.

\section{A look into the future}
Let us illustrate how the current picture in Fig.\ \ref{fig:Autumn24_ext} might evolve towards the end of the Belle-II and HL-LHC physics programmes.
We therefore recalculate the central values for the CP asymmetry observables to correspond to the solution obtained from the extended fit, i.e.\ Eqs.\ \eqref{eq:extended_fit_I}--\eqref{eq:extended_fit_IV}.
The expected uncertainties on the input observables are taken from the prospects published by LHCb \cite{LHCb:2018roe} and Belle-II \cite{Belle-II:2018jsg} where possible.
Otherwise, we extrapolate the current statistical uncertainty with the expected increase in luminosity, and conservatively assume the current systematic uncertainty to be fully irreducible.

\begin{figure}
    \centering
    \includegraphics[width=0.47\textwidth]{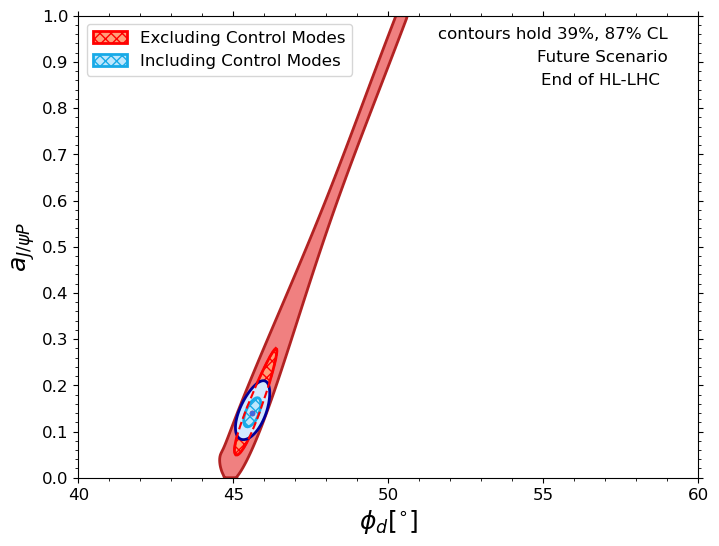}
    \includegraphics[width=0.47\textwidth]{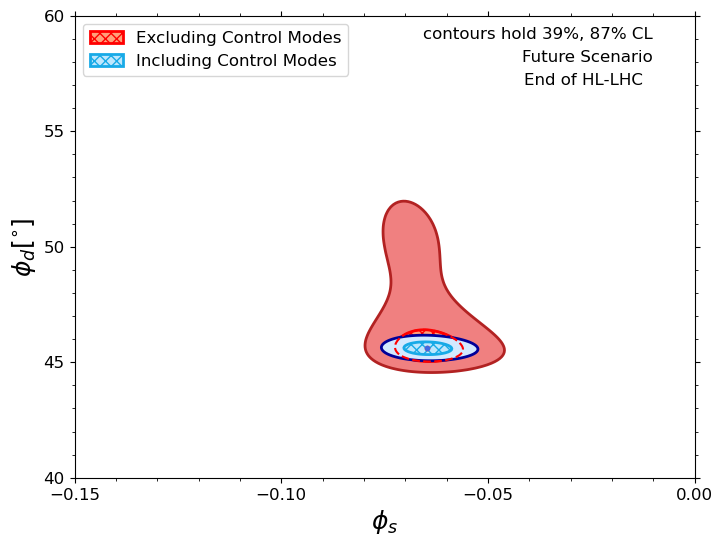}
    \caption{Two-dimensional confidence regions of the fit for the penguin parameters, $\phi_d$ and $\phi_s$ from potential future CP asymmetry measurements in the $B\to J/\psi X$ and $B\to DD$ decays.
    Comparison between the two future scenarios for the expected situation after the end of the Belle-II and HL-LHC programmes.
    }
    \label{fig:Future_phi}
\end{figure}

In Fig.\ \ref{fig:Future_phi}, we compare the two-dimensional confidence regions for two future scenarios:
\begin{enumerate}
    \item \textbf{Excluding control modes:} In this scenario we only consider updated measurements for the modes $B_d^0\to J/\psi K_{\mathrm{S}}^0$, $B_s^0\to J/\psi\phi$ and $B_s^0\to D_s^+D_s^-$.
    We assume no new CP asymmetry measurements will become available for the penguin control modes $B_d^0\to J/\psi\pi^0$, \mbox{$B_s^0\to J/\psi K_{\text{S}}^0$,} $B_s^0\to J/\psi \rho^0$ and $B_d^0\to D^+D^-$.
    \item \textbf{Including control modes:} In this scenario we assume the seven decay channels analysed in this paper receive equal attention and new CP asymmetry measurements will become available for all of them.
\end{enumerate}

In the first scenario we expect an improvement in the determination of $\phi_d$ by approximately 30\% compared to today.
This can be increased by another factor two if also the penguin control modes are updated with the full data sets available (second scenario).
Regarding the determination of $\phi_s$, the increased luminosity provided by the HL-LHC will reduce the uncertainty by at least a factor two compared to today (first scenario).
Updating the penguin control modes (second scenario) will provide another 15\% to 30\% improvement.
An improved determination of the penguin shift $\Delta\phi_q$ can thus lead to significant gains in the expected precision on $\phi_d$ and $\phi_s$, which is necessary to maximally benefit from the Belle-II and HL-LHC programmes in the search for BSM physics.

\section{Conclusion}
These proceedings present the latest results from a data-driven approach, utilizing the $SU(3)$ flavour symmetry of QCD, to control the penguin contributions in the determination of $\phi_d$ and $\phi_s$ from the $B_d^0\to J/\psi K_{\mathrm{S}}^0$, $B_s^0\to J/\psi\phi$ and $B_s^0\to D_s^+D_s^-$ decays.
Using the current CP asymmetry measurements from these three decay channels and their control modes we find
\begin{align}
    \phi_s & = -0.065_{-0.017}^{+0.019}\:\text{(exp)} \pm 0.002\:(SU(3))
    = \left[-3.72_{-0.97}^{+1.09}\:\text{(exp)} \pm 0.11\:(SU(3))\right]^{\circ}\:,\\
    \phi_d & = \left[45.6_{-1.0}^{+1.1}\:\text{(exp)} \pm 0.3\:(SU(3))\right]^{\circ}\:,
\end{align}
where the first uncertainty comes from our fit to the experimental inputs and the second uncertainty estimates the potential impact from $SU(3)$-breaking effects.

To reach a precision on $\phi_s$ below 10 mrad $\left(0.6^{\circ}\right)$ and/or on $\phi_d$ below $1^{\circ}$, as anticipated for the end of Belle-II and HL-LHC physics programmes, we will need to improve our understanding of the penguin effects through improved measurements of the control modes.

\section*{Acknowledgements}
This research has been supported by the Netherlands Organisation for Scientific Research (NWO), and the German Research Foundation (DFG) under grant 396021762 - TRR 257. 

\phantomsection 
\addcontentsline{toc}{section}{References}
\setboolean{inbibliography}{true}
\bibliographystyle{LHCb}
{\small \bibliography{references}}

\end{document}